\documentclass[apjl]{emulateapj}
\usepackage{epsf}
\usepackage{amssymb,amsmath}
\usepackage{multirow}
\usepackage{rotating}
\usepackage{subfigure}
\usepackage{color}

\newcommand\resetsubfigs{\setcounter{sub\@captype}{0}}

\def\apj{ApJ}

\def\apjl{ApJL}
\def\apjs{ApJS}
\def\mnras{MNRAS}

\def\pasp{PASP}

\def\nat{Nat}

\def\apjs{ApJS}

\def\MSUN{M$_{\sun}$ }
\def\kms{$\rm \ km~s^{-1}$ }
\def\Mdot{\.{M}}

\begin{document}
\slugcomment{Submitted to ApJ Letters on August 5, 2015}
\title{High Resolution {\it Chandra} HETG spectroscopy of V404  Cygni in Outburst}
\author{Ashley L. King\altaffilmark{1,2}, Jon M. Miller\altaffilmark{3},  John Raymond\altaffilmark{4}, Mark T. Reynolds\altaffilmark{3}, Warren Morningstar\altaffilmark{1}  }
\altaffiltext{1}{Department of Physics, 382 Via Pueblo Mall, Stanford, CA 94305, ashking@stanford.edu}
\altaffiltext{2}{Einstein Fellow}
\altaffiltext{3}{Department of Astronomy, University of Michigan, 1085 S. University Ave, Ann Arbor, MI 48109-1107, USA}
\altaffiltext{4}{Harvard-Smithsonian Center for Astrophysics, 60 Garden St., Cambridge, MA 02138, USA}

\begin{abstract}
As one of the best-characterized stellar-mass black holes, with good
measurements of its mass, distance and inclination, V404 Cyg is the
ideal candidate to study Eddington-limited accretion episodes.  After
a long quiescent period, V404 Cyg underwent a new outburst in June
2015.  We obtained two {\it Chandra} HETG exposures of 20 ksec and 25
ksec.  Many strong emission lines are observed; the ratio of Si
He-like triplet lines gives an estimate for the formation region distance of $4\times10^{11}$ cm, while the higher ionization Fe XXV He-like triplet gives an estimate of $7\times10^9$ cm.   A narrow Fe K$\alpha$ line is detected
with an equivalent width greater than 1 keV in many epochs, signaling
that we do not directly observe the central engine.  Obscuration of
the central engine and strong narrow emission lines signal that the
outer disk may be illuminated, and its structure may help to drive the
strong variability observed in V404 Cyg.  In the highest flux phases,
strong P-Cygni profiles consistent with a strong disk wind are
observed.  The kinetic power of this wind may be extremely high.
\end{abstract}
\maketitle

\section{Introduction}
Stellar-mass black holes are known for extreme behaviors, but V404 Cyg
stands out even within this class. V404 Cyg, also known as GS 2023$+$338,  was discovered with the all-sky monitor aboard {\it Ginga} on
1989 May 22 \citep{Makino89}. The transient was extremely bright and
observed extensively, varying by a factor of $\sim500$ on timescales of seconds \citep{Kitamoto89}.  A low-mass
companion was found in a wide, 6.5-day binary, and dynamical
constraints verified that V404 Cyg indeed harbors a black hole of
$9.0^{+0.2}_{-0.6}$ \MSUN with a binary inclination of $(67^{+3}_{-1})^\circ$ \citep{Casares92,Sanwal96}.  Recent radio parallax measurements give a distance of
$2.39\pm 0.14$~kpc \citep{Miller-Jones09}. The Eddington luminosity for V404 Cyg is
$1.1\times10^{39}$ ergs s$^{-1}$

Observations of disk winds in stellar-mass black holes in the {\it
  Chandra} era find them to be dense, equatorial, to originate as
close as $\sim 1000~ GM/c^{2}$ \citep{Miller06,Miller08}, and to be
anti-correlated with jets \citep{Miller06b, Miller08,Neilsen09,King12, Ponti12}.  Winds may be
related to the basic physics of disk accretion, and offer a view into
feedback between massive black hole and host galaxies \citep[e.g.][]{King13}. 
 New results find P-Cygni profiles and evidence of
Keplerian rotation in the wind, and that components with velocities of
$v/c \simeq 0.01$ may be fairly common \citep{Miller15}.
However, although mass outflow rates may exceed the accretion rate at
the inner disk, the kinetic power in winds is (so far) a small
fraction of the radiated power \citep{King13}.  In this sense,
even the most extreme stellar-mass black hole winds \citep[with velocities
approaching 0.03-0.05$c$;][]{King12}, differ from the powerful
outflows observed in broad absorption line quasars (BALQSOs) and
active galactic nuclei (AGN) such like PDS 456 \citep{Nardini15}.

A new outburst of V404 Cyg was discovered using the {\it Swift}/BAT on
2015 June 15 \citep{Barthelmy15}.  Rapid, extreme variability was
again observed; flaring up to several tens of Crab was observed with
{\it INTEGRAL} \citep{Natalucci15,Rodriguez15}.  Remarkably, this does not
automatically correspond to a super-Eddington luminosity, but rather a
luminosity of $L \simeq 4.3\times10^{38}~ {\rm erg}~ {\rm s}^{-1}
\simeq 0.4~L_{Edd}$.  Motivated by the opportunity to understand the
origin of the extreme behaviors observed from V404 Cyg in terms of its
accretion inflow and outflows, we obtained an observation using the
{\it Chandra} High Energy Transmission Grating Spectrometer \citep[HETGS;][]{Canizares05}.  
Initial results from our analysis of the
incredibly rich spectra that were obtained by \cite{King15} are the
focus of this Letter.

\section{Methods}
We obtained two observations with the {\it Chandra} High Energy
Transition Grating Spectrometer (HETGS) on 22 June 2015 and 23 June
2015, for 21 ksec and 25 ksec, respectively.  Owing to the
extraordinary flux levels exhibited by the source, the observations
were taken with the ACIS-S array in continuous clocking mode with the 0th order position located off the detector array.  This
was a necessary precaution needed to avoid
radiation damage to the ACIS CCDs.  Used once before to observe Sco
X-1, this is an unusual instrumental configuration that requires
careful reduction procedures.

The resulting spectra from our setup include the positive orders of
the medium energy grating (MEG) and the negative orders of the high
energy grating (HEG). We used the {\tt ciao} tools, version 4.7, to
reduce the data. The pha files were extracted with {\tt tgextract}
using a masked region that was centered on the nominal point source
position (RA 20h 24m 3.834s, DEC 33$^\circ$ 52$'$ 2.33$''$).  This
position was then manually iterated in successive reductions, so that
the narrow Fe K$\alpha$ line measured centroid energy (as measured
with a Gaussian) was aligned in both the first and third MEG positive
orders as well as the HEG first negative orders, to within measurement
errors.

The response files were created with {\tt mktgresp}.  Prominent dips
in the high energy of the MEG light curve were observed to coincide
with the dither pattern period (707 seconds).  We therefore excluded
the last 10 rows of the masked files to remove this periodic behavior.
Periodic steepening of the spectra owing to this effect was thereby
eliminated.

The configuration also impacted the useful energy range in each
dispersed spectrum.  We analyzed data from the MEG between 1.4--6 keV,
and from the HEG between 5.6--10 keV in the first observation.  In the
second observation, we utilized data from the MEG between 1.4--4 keV,
and from the HEG between 3.5--10 keV.  The selected energy ranges vary
between the observations due to differences in the effective area
curves resulting from the specific pointings, and strong backgrounds
observed in the HEG.  In a follow-up paper, we will perform a more
in-depth analysis of this background with an aim at extending the
bandpass to lower energies in order to study Ne, Mg, and Fe L-shell
lines.

All spectra were grouped to require at least 10 counts per bin, as per
Cash (1979).  Grouping only improved the fits above 7~keV.

\section{Analysis}
\subsection{Time-Averaged Spectra}
The time-averaged spectra for the two separate observations are
depicted in Figure 1a\&b. The observed 2--10 fluxes of these
observations are 9.5$\times10^{-9}$ ergs s$^{-1}$ cm$^{-2}$ and
1.3$\times10^{-8}$ ergs s$^{-1}$ cm$^{-2}$, respectively. This
corresponds to an observed luminosity of 6.5$\times10^{36}$ ergs
s$^{-1}$ and 8.9$\times10^{36}$ ergs s$^{-1}$, respectively, and does
not take into account any absorption along our line-of-sight.

We initially fit the broadband continuum with a phenomenological
power-law and a broad Gaussian line (restricted to the 5--7 keV
range), as a relativistically broadened Fe K$\alpha$ line is likely present (See Figure 1a
\& 1b).  Narrow Gaussians lines were then fit to Mg XII, Si XIII,
Si XIV, SXV, SXVI, Fe K$\alpha$, Fe XXV, Fe XVVI, and Fe K$\beta$ (See Table 1). The energy,
line width, and normalizations of these lines were allowed to vary.
However, the energies and line widths of the forbidden (f) and
inter-combination (i) lines were tied to each other.

The lower ionization lines are generally slightly red-shifted at 200\kms, with FWHM
at roughly 1700\kms. The higher ionizations lines, especially Fe XXVI are blue shifted at $-$1000\kms with FWHM line widths of 3000\kms. Finally, the neutral Fe K$\alpha$ line is at line center with a FWHM of 3400\kms, while the Fe K$\beta$ is redshifted by 700\kms with a FWHM of 2500\kms.

\subsection{Time-Resolved Spectra}

After examining the time-averaged spectra, we divided each observation
into 3.2 ksec segments. This afforded a  minimum
of 13400 counts in the even the lowest flux segment, while still
enabling us to track the large scale variations in both the flux and
spectral hardness.

Figures 2a--h plot the ratio of each spectrum to a simple
power-law. Both the spectral index, $\Gamma$, and normalization were
allowed to vary in these fits. One can see changes in the ratios
of the lines to the continuum and the ratio of 
the neutral Fe K$\alpha$ \&$\beta$ lines to Fe XXV \& XXVI, indicating changes in
the highest ionization ions.  In addition, several observations show clear absorption features (the fourth panel from the bottom in Figure 2 b and the top two panels of Figure 2 e--h).

Measuring the line strength of the narrow Fe K$\alpha$, which is the
line with the highest signal to noise ratio, we find that the line
strength is positively correlated with the observed flux (Figure
3a). A positive correlation may indicate the line is responding to the
continuum.  In contrast, the equivalent width of the line is inversely
correlated with the observed flux, and exceeds 1 keV at the lowest
continuum fluxes (Figure 3b). An inverse correlation between
equivalent width and observed flux suggests the continuum is varying
independently of a relatively constant Fe K$\alpha$ line, thus
illustrating the complexity of this system.  The fact that the
equivalent width of the Fe K$\alpha$ line sometimes exceeds 1 keV
signals that we are likely not viewing the direct continuum that is
photoexciting the Fe K$\alpha$ line, but rather a reflected or
scattered flux into our line of sight. For reference an equivalent width of 0.1 keV is expected when directly viewing the central source \citep{George91}. A high equivalent width is consistent with the low
spectral indices $\Gamma<1.4$ for most of the observations (Figure
3c).  Reflection may play an important role in these spectra.  In a
follow-up paper, we will further discuss the continuum variations.

Finally, in addition to emission features, there are clear absorption
features in several of the brightest epochs.  Figure 4a shows the
absorption in the 8th epoch of the second observation.  A P-Cygni
profile is observed, with absorption blue-ward of the red-shifted
emission lines. This type of line profile is detected in all but the
neutral Fe K$\alpha$ and Fe K$\beta$ lines, which likely originate at
a larger radius than the other ions. The absorption features broaden
from 1000\kms in the Mg XII line, to nearly 4000\kms in the Fe XXVI
line.  There is clearly  a complex structure to the wind that is
generated during this outburst.  For contrast, line profiles of
the 6th epoch are shown in Figure 3b: lines from that epoch are
narrow, close to their rest energy values, and seen only in emission.

\subsection{Emission and Absorption Lines}

Three sorts of narrow spectral lines are observed.  First, there are
forbidden and intercombination lines (f and i) of He-like ions formed
by recombination in photoionized gas.  Resonance lines of He-like ions
(labeled r) and Lyman lines of H-like ions can be formed in the same
way, but recombination produces r lines that are only about 1/3 as
strong as the sum of f plus i.  Second, P-Cygni profiles with blue
edges corresponding to $V_{wind}$ reaching 4000\kms are apparent in the
time-resolved spectra.  While a spherical wind gives P-Cygni emission
weaker than absorption, a wind from a disk seen at high inclination
can give emission much stronger than absorption, as is seen in high
\Mdot \ cataclysmic variables \citep{Mauche87}.  Third, the Fe
K$\alpha$ and K$\beta$ lines are formed by inner shell photoionization
of weakly ionized iron in relatively cold gas.

The ratio of f to i lines is a density diagnostic,
but in the presence of a strong UV radiation field the ratio is
determined by photoexcitation between the upper levels of those lines
\citep{Mauche03}.  The Si XIII f to i ratio in the first observation
of 0.9 would indicate a density of about $3\times
  10^{13}{\rm cm}^{-3}$, while the Fe XXV f to i ratio of the first observation of 0.5, would indicate a density of $10^{17}$ cm$^{-3}$. Of course this assumes the UV radiation field is unimportant.  However, the peak V-band magnitude of about {11.3} from AAVSO observations that day together with $\rm
A_V$ = 3.0 \citep{Shahbaz96} and an $\rm F_\nu \propto \nu^{1/3}$
accretion disk spectrum imply a UV continuum luminosity that gives an Si XIII f to i ratio of 0.9 at
 $\sim4\times10^{11}$ cm from the UV source, and a Fe XXV f to i ratio of 0.5 at $\sim 7\times10^9$ cm. The apparent brightness should be corrected for the 70$^\circ$ inclination of the
disk and the obscuration of part of the disk by its edge, suggesting
that the actual UV luminosity is an order of magnitude larger, and the
diffuse, emitting gas is located at about 10$^{12}$ cm and $10^{10}$ cm from the
source for the Si XII and Fe XXV, respectively. Interestingly this points to two different regions separated by nearly two orders of magnitude in distance, consistent the wide range in ionization between these two ions. 

The ratio of He-like to H-like lines serves as an indicator of the
ionization parameter, $\xi \equiv L/n r^2$.  If the lines are
primarily formed by recombination, the approximate equality of the
intensities of the He-like complex and H-like Ly$\alpha$ of silicon
implies approximately equal amounts of H-like S XIV and ion Si XV,
which in turn implies log $\xi \approx$ 3.0 \citep{Kallman82}.
The presence of Fe XXV and Fe XXVI implies a higher ionization
parameter, log $\xi$ at least 4.0, while the presence of P-Cygni
absorption in the Si XIII r line in some time-resolved spectra implies
log $\xi$ below 3.  There is likely a continuous range of
ionization parameters in the emitting gas.

The f and i lines of Si XIII are formed by recombination of Si XIV to
the excited triplet states of Si XIII. Their observed
intensities, a distance of 2.39 kpc \citep{Miller-Jones09} and a factor of 1.4
correction for interstellar absorption, give an emission measure (EM) of
$2\times10^{58} {\rm cm}^{-3}$. For a $1/r^2$ density distribution, $N_H = nr$, and we can
roughly approximate the emission measure as $EM=n^2r^3$ under the
assumption that much of the emitting volume is obscured by the
disk. Then
\begin{equation}
r=\frac{nr^2}{nr}=\frac{L/\xi}{N_H}=\frac{n^2r^3}{(nr)^2}=\frac{EM}{N_H^2} 
\end{equation}
\noindent
yields $N_H = 10^{23}$ cm$^{-2}$, $r = 2\times10^{12}$ cm, and $n
= 5\times10^{10}$ cm$^{-3}$.  These are fairly crude estimates
considering the enormous variability in both the lines and the
continuum, but they are consistent with the diagnostics given above. 
\subsection{Wind parameters}

As a starting point, we assume that all the lines are formed in wind.  Then

\begin{equation}
\dot{M}= 4 \pi \frac{\Omega}{4 \pi} \mu n r^2 V_{wind} = 4 \pi \frac{\Omega}{4 \pi} \frac{\mu L}{\xi}V_{wind}
\end{equation}

\noindent
and

\begin{equation}
L_{wind} = \frac{1}{2} \dot{M} V_{wind}^2 = 2 \pi \frac{\Omega}{4 \pi} \frac{ \mu L}{\xi}V_{wind}^3
\end{equation}

\noindent
or

\begin{equation}
\frac{L_{wind}}{L} = 2 \pi \frac{\Omega}{4 \pi \xi} \mu V_{wind}^3   ,
\end{equation}

\noindent
where $\Omega$ is the covering fraction of the wind and $\mu$ is the
mean atomic mass.  While $\Omega$ is on the order of 0.1 in GRO
J1655-40 \citep{Miller06}, the emission lines and the strong
absorption at a lower inclination indicate that $\Omega$ is closer to
1 in V404 Cyg.  Thus the ionization parameter and $V_{wind}$ determined above
imply $L_{wind}/L$ is about {\bf 0.1}. V404 Cyg may be reaching its Eddington
limit, providing enough radiation pressure to disrupt the outer disk,
generating these features.  It is interesting to note that this is
comparable to the level of feedback required in some simulations to
explain the well-known $M-\sigma$ relationship \citep{DiMatteo05}.

We can also estimate the column density of the wind.  The P-Cygni
absorption of Si XIV Ly$\alpha$ extends beyond 4000\kms in some
spectra.  The optical depth is about 0.1 at high speeds, which
requires a Si XIV column density of about $7\times10^{16}~\rm
cm^{-2}$ and a hydrogen column of $4\times10^{22} ~\rm cm^{-2}$ if
half the silicon is Si XIV. The lack of a clear Si XIV absorption edge
at 2.67 keV implies a limit to its optical depth of $\tau_{Si}< 0.2$,
giving $N_H <4\times10^{23} {\rm cm}^{-2}$, in agreement with the column density estimate in the previous section.

\subsection{Line Origins}

While the above is a reasonable interpretation of the emission and
absorption features, the wild variability of the line strengths
(Figure 3) means that it is (at best) some sort of average
description of the wind, and alternate interpretations seem quite
possible.

The most basic question is whether the emission lines (in particular
the f and i lines of He-like ions) are formed in the wind where the P
Cygni profiles form.  The f and i lines are relatively narrow, 1700\kms FWHM,
compared with $V_{\rm wind} > 4000$\kms from the blue edge of the P
Cygni profile and the FWHM=3400\kms higher ionization Fe XXVI lines.  The narrow width is compatible with Keplerian rotation at
the radius of $2\times10^{11}$ cm (3.1$\times10^{5}\ GM/c^2$)
inferred above, considering that the narrow peak requires emission from larger radii.
This is a plausible size for the outer edge of the disk.

A possibility that would explain the relatively narrow line widths,
would be that the Si and S f and i lines arise in the X-ray-illuminated outer disk, as in Her X-1 in
the low state \citep{Jimenez05}, while the P-Cygni lines form in a
more or less unrelated wind.  To match the line-to-continuum ratios
and equivalent widths, especially in the fainter spectra in Figures
 2 and 3, this picture requires that the outer edge of the disk
block our line of sight to the central X-ray source, so that we
observe continuum produced by scattering from the disk or in an
Accretion Disk Corona (ADC).  In that case $L_X$, which already
appears to be up to {\bf one tenth $L_{edd}$}, must be even higher. This means that the observed mass accretion rate \Mdot$_{acc}\approx 10^{19}\ $g~s$^{-1}$ (assuming 0.1 efficiency) is also likely much higher.
In this picture, the continuum variability, which is stronger than the
variability in the emission lines (Figures 2 a--h \& 3) could be attributed to
changes in the height of the edge of the disk rather than changes in
accretion rate. This is also evidenced by the changing spectral
hardness (Figure 3c), which suggests a variable absorption and/or
reflection contribution to the continuum.

\section{Summary}
The two {\it Chandra} HETG observations taken on the 22 and 23 of June
2015, show strong spectral emission and absorption features. Mg XII, Si
XIII, Si XIV, S XV, S XVI, Fe K$\alpha$ \& K$\beta$, Fe XXV, and Fe
XXVI are detected in the time-averaged spectra, as
well as the time-resolved spectra. Variability of the line strengths compared to the continuum
as well as changes in Fe K$\alpha$ to Fe XXV ratios
suggests changes in equivalent widths as well as the highest ionizations during the
observations. In addition, the extreme equivalent widths of the Fe
K$\alpha$ line (EW$>1$ keV) and continuum spectral hardness, indicates
that we are not directly observing the ionizing continuum, and
that the total luminosity can exceed the $L_{2-10\ keV }=0.1 L_{Edd}$ apparent
luminosity.

The ratio of the Si XIII f to i lines gives a distance from the central source
of $4\times10^{11}$ cm, while the ratio of the Fe XXV f to i lines gives a distance of $7\times10^{9}$ cm.  The emission measure, luminosity and ionization parameter of Si XIII
yields a somewhat larger distance, along with a density of $5\times10^{10}$ cm$^{-3}$.
The low ionization density estimate is several orders of magnitude smaller than the density seen in an outburst of GRO1655-40 \citep{Miller06}, presumably because the gas is much farther from the central
source. In addition, the winds from V404 Cyg
could be associated with disruption of the outer disk from the
radiation pressure of the central region as it reaches or exceeds the
Eddington luminosity. In contrast, radiation pressure is negligible in typical X-ray binary winds \citep{Miller06,King13}.

In a follow-up paper, we will further discuss the continuum, quantifying the amount of reprocessed scattered or reflected light that is directed into our line of sight. In addition, we will include a physically motivated, photo-ionization model to characterize the ions observed in both emission and absorption described in this paper.

\section*{Acknowledgements}
The authors would like to thank thank Belinda Wilkes for this DDT, as well as Herman Marshall, David Huenemoerder, and the {\it Chandra} team for their invaluable help. ALK would like to thank the support for this work, which was provided by NASA through Einstein Postdoctoral Fellowship grant number PF4-150125 awarded by the Chandra X-ray Center, operated by the Smithsonian Astrophysical Observatory for NASA under contract NAS8-03060.

\begin{deluxetable*}{lllllll} 
\tabletypesize{\small}
\tablecolumns{4} 
\tablewidth{0pc} 
\tablecaption{Model Parameters} 
\tablehead{ 
\colhead{}  & \colhead{Observation 1} & \colhead{Observation 2} & \colhead{E$_{0}$ (keV)} & \colhead{Line ID}}
\startdata
Energy (keV) & 1.4725$^{+0.0002}_{-0.0003}$ & 1.4715$^{+0.0002}_{-0.0001}$ & 1.472  & Mg XII (Lyman $\alpha$)\\
$\sigma$ (eV) &   3.7$\pm{0.3}$ &  2.0$^{+  0.2}_{-  0.1}$&  \\
Norm ($10^{-3}$ photons cm$^{-2}$ s$^{-1}$) &   1.84$^{+    0.06}_{-  0.12}$  & 2.4$^{+  0.2}_{-  0.1}$ & \\
\\
Energy (keV) & 1.8411* & 1.8401*  & 1.839 & Si XIII (f)\\
$\sigma$ (eV) &  3.0* &  2.8*\\
$N_{R}$ ($10^{-3}$ photons cm$^{-2}$ s$^{-1}$) &   0.79$^{+    0.07}_{-  0.06}$  &   1.7$\pm0.1$  \\
\\
Energy (keV) & 1.8557$^{+0.0003}_{-0.0004}$  & 1.8547$^{+0.0002}_{-0.0003}$  & 1.854 & Si XIII (i)\\
$\sigma$ (eV) &   3.0$^{+    0.3}_{-  0.2}$ &   2.8$\pm 0.2$ \\
$N_{I}$ ($10^{-3}$ photons cm$^{-2}$ s$^{-1}$)  &     0.84$\pm{-0.07}$ &   1.6$\pm0.1$  \\
\\
Energy (keV) & 1.8651$^{+0.0003}_{-0.0006}$  & 1.8642$^{+0.0003}_{-0.0003}$    &  1.865 & Si XIII (r)\\
$\sigma$ (eV) &    2.3$^{+    0.5}_{-  0.4}$  & 1.8$^{+  0.2}_{-  0.4}$  \\
$N_{F}$ ($10^{-3}$ photons cm$^{-2}$ s$^{-1}$)  &     0.63$^{+    0.06}_{-  0.07}$   &   1.2$\pm0.1$  & \\
\\
Energy (keV) &  2.0050$^{+0.0003}_{-0.0002}$  & 2.0042$^{+0.0002}_{-0.0002}$ & 2.005 & Si XIV (Lyman $\alpha$)\\
$\sigma$ (eV) &  3.4$^{+    0.3}_{-  0.2}$ &   2.7$\pm0.2$  & \\
Norm ($10^{-3}$ photons cm$^{-2}$ s$^{-1}$)  &  2.05$^{+    0.08}_{-  0.09}$  &   3.7$^{+  0.1}_{-  0.2}$ &  \\
\\
Energy (keV) & 2.4336* & 2.4362* & 2.430 &  S XV (f)\\
$\sigma$ (eV) &   3.5* &  6* \\
$N_{r}$ ($10^{-3}$ photons cm$^{-2}$ s$^{-1}$)  &   0.8$^{+    0.2}_{-  0.1}$ &   2.1$^{+  0.2}_{-  0.3}$ \\
\\
Energy (keV) & 2.4513$^{+0.0008}_{-0.0010}$ & 2.454$^{+0.001}_{-0.001}$  & 2.447 & S XV (i)\\
$\sigma$ (eV) &  3.5$^{+    1.0}_{-  0.7}$ &   6$^{+  1}_{- 1}$  & \\
$N_{i}$ ($10^{-3}$ photons cm$^{-2}$ s$^{-1}$) &  0.9$^{+    0.2}_{-  0.1}$  &   1.3$^{+  0.2}_{-  0.4}$ \\
\\
Energy (keV) & 2.4625$^{+0.0049}_{-0.0001}$  & 2.461$^{+0.002}_{-0.001}$ & 2.461 &  S XV  (r)\\
$\sigma$ (eV) &  0.07$^{+    3.05}_{-  0.07}$ &   1$^{+  2}_{-  1}$   \\
$N_{f}$ ($10^{-3}$ photons cm$^{-2}$ s$^{-1}$)  &  0.4$\pm{  0.1}$ &   0.7$^{+  0.1}_{-  0.3}$ \\
 \\
Energy (keV) & 2.4336* & 2.4362* & 2.430 &  S XV (f)\\
$\sigma$ (eV) &   3.5* &  6* \\
$N_{r}$ ($10^{-3}$ photons cm$^{-2}$ s$^{-1}$)  &   0.8$^{+    0.2}_{-  0.1}$ &   2.1$^{+  0.2}_{-  0.3}$ \\
\\
Energy (keV) & 2.620$\pm{0.0010}$  & 2.6197$^{+0.0006}_{-0.0006}$  & 2.620 & S XVI (Lyman $\alpha$)\\
$\sigma$ (eV) &  4.3$^{+    0.8}_{-  1.2}$&   4.8$\pm0.7$  \\
Norm ($10^{-3}$ photons cm$^{-2}$ s$^{-1}$)  &  1.1$^{+    0.1}_{-  0.2}$ &   3.6$\pm0.3$  \\
\\
Energy (keV) & 6.395$\pm0.001$  & 6.3970$^{+0.0004}_{-0.0017}$  & 6.403/6.391 &  Fe K$\alpha_1$,$\alpha_2$  \\
$\sigma$ (eV) &  21$^{+    2}_{-  1}$     &  18$^{+  1}_{- 1}$  \\
$N_{f}$ ($10^{-3}$ photons cm$^{-2}$ s$^{-1}$)  &  24.7$^{+    0.8}_{-  1.0}$ &  28.1$^{+  0.9}_{-  0.8}$   \\
\\
Energy (keV) & 6.613* & 6.613* & 6.637 & Fe XXV (f) \\
$\sigma$ (eV) &  16* & 16*\\
Norm  ($10^{-3}$ photons cm$^{-2}$ s$^{-1}$) &      1.5$^{+  0.8}_{-  0.4}$ & 1.8$^{+    0.4}_{-  0.6}$  &  \\
\\
Energy (keV) & 6.651*   & 6.651* & 6.682 & Fe XXV (i) \\
$\sigma$ (eV) & 16* & 16*  \\
Norm  ($10^{-3}$ photons cm$^{-2}$ s$^{-1}$) &   2.8$^{+  0.9}_{-  0.6}$ &   2.8$^{+    0.4}_{-  0.8}$    \\
\\
Energy (keV) & 6.677*   &  6.677*  & 6.700 & Fe XXV (r) \\
$\sigma$ (eV) &  16 * & 16* \\
Norm  ($10^{-3}$ photons cm$^{-2}$ s$^{-1}$)  &   3.6$^{+  0.5}_{-  1.0}$ &   3.2$^{+    0.3}_{- 1.0}$   \\
\\
Energy (keV) & 6.948* & 6.948$^{+0.003}_{-0.003}$ & 6.973 & Fe XXVI (Lyman $\alpha$)\\
$\sigma$ (eV) &  17*  &  17$^{+  5}_{-  3}$  \\
$Norm$ ($10^{-3}$ photons cm$^{-2}$ s$^{-1}$)  &    5.3$^{+    0.6}_{-  1.1}$   & 7.2$^{+  0.6}_{-  0.8}$ \\
\\
Energy (keV) & 7.074*  & 7.074$^{+0.004}_{-0.006}$  & 7.058 &  Fe K$\beta$\\
$\sigma$ (eV) &  14*  &  14$^{+  7}_{- 8}$  \\
Norm  ($10^{-3}$ photons cm$^{-2}$ s$^{-1}$) &  3.9$^{+    0.5}_{-  1.1}$  &   4.1$^{+  0.6}_{-  0.8}$ \\
\\

\enddata
\tablecomments{These are the best fit parameters to the emission lines in both observations. The * refers to the components that are frozen.}
\end{deluxetable*}

\begin{figure*}
\subfigure[\label{fig:1}][]{\includegraphics[width=.9\linewidth,angle=0]{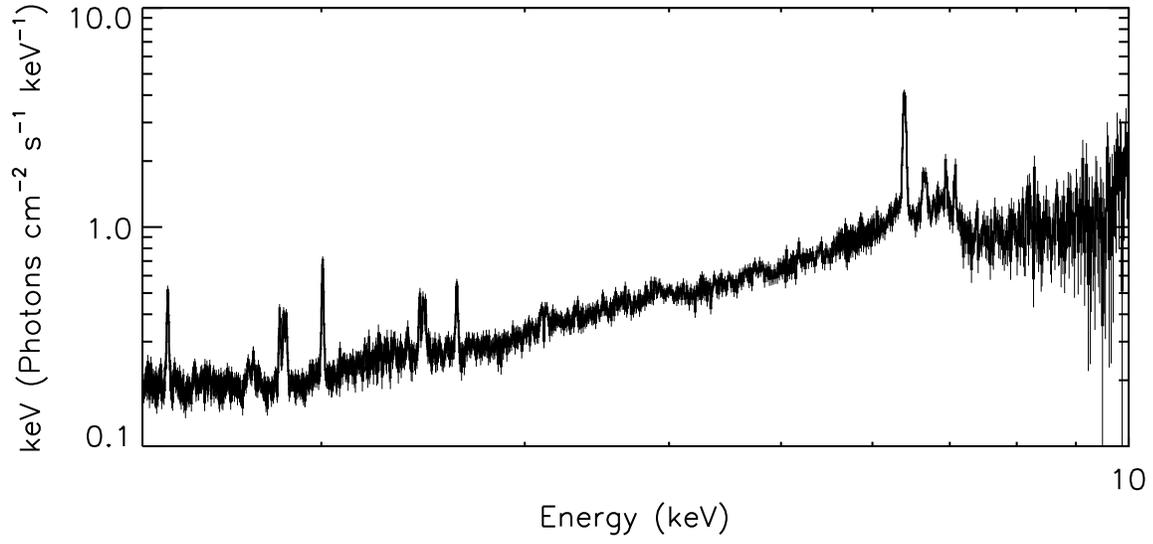}}

\subfigure[\label{fig:1}][]{\includegraphics[width=.9\linewidth,angle=0]{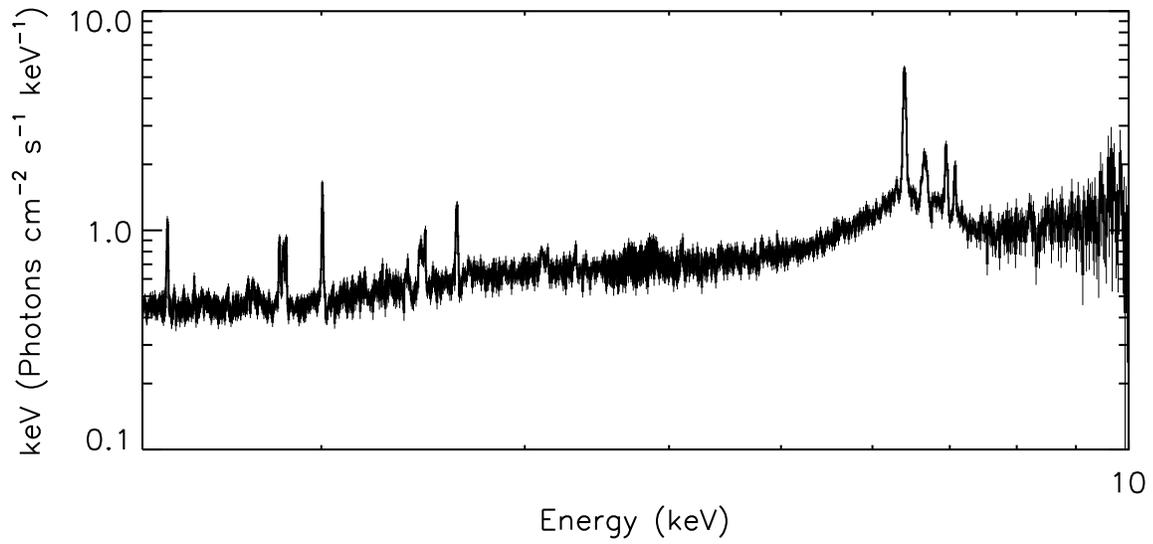}}

\caption{ This figure shows the unfolded spectrum of the first (A) and second (B) observations. Emission features are observed in Mg XII, Si XIII, Si XIV, S XV, S XVI, Fe K$\alpha$ \& K$\beta$, Fe XXV, and Fe
XXVI. In addition a broad excess is observed around 6.4 keV, but has a slightly different shape in the two observations.  \label{fig:obs1} }
\end{figure*}

\begin{figure*}
\subfigure[\label{fig:1}][]{\includegraphics[width=.23\linewidth,angle=0,clip=true,trim=0cm 2cm 0   6cm]{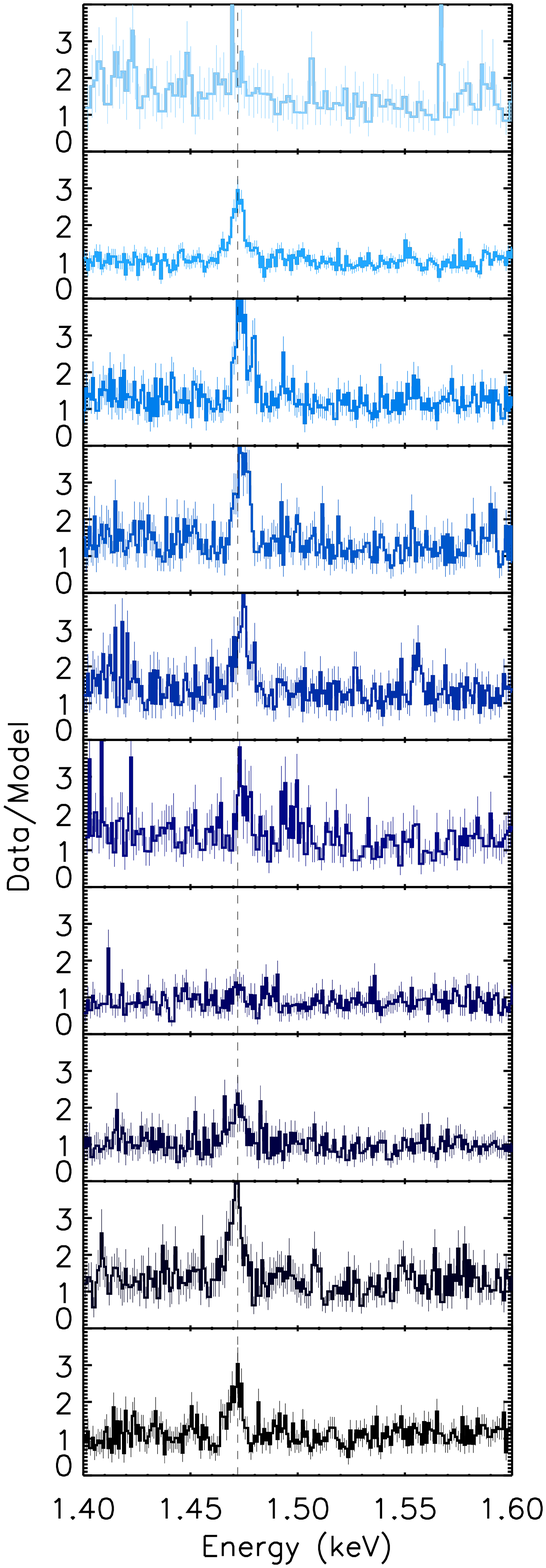}}
\subfigure[\label{fig:1}][]{\includegraphics[width=.23\linewidth,angle=0,clip=true,trim=0cm 2cm  0   6cm]{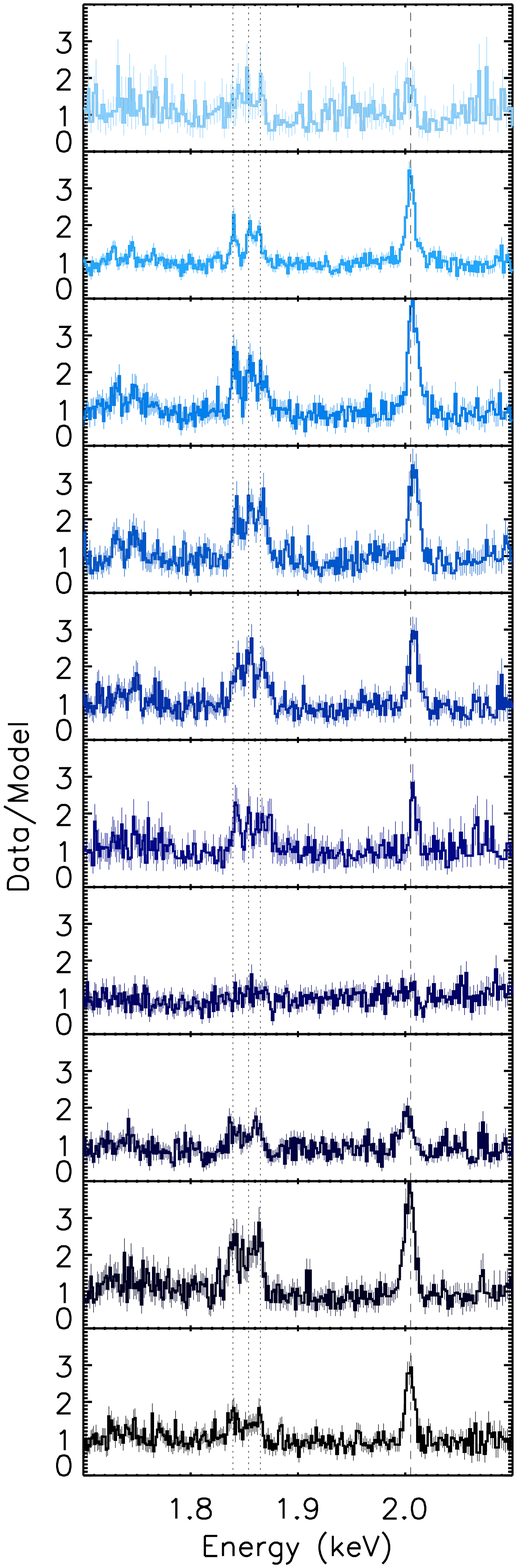}}
\subfigure[\label{fig:1}][]{\includegraphics[width=.23\linewidth,angle=0,clip=true,trim=0cm 2cm 0  6cm]{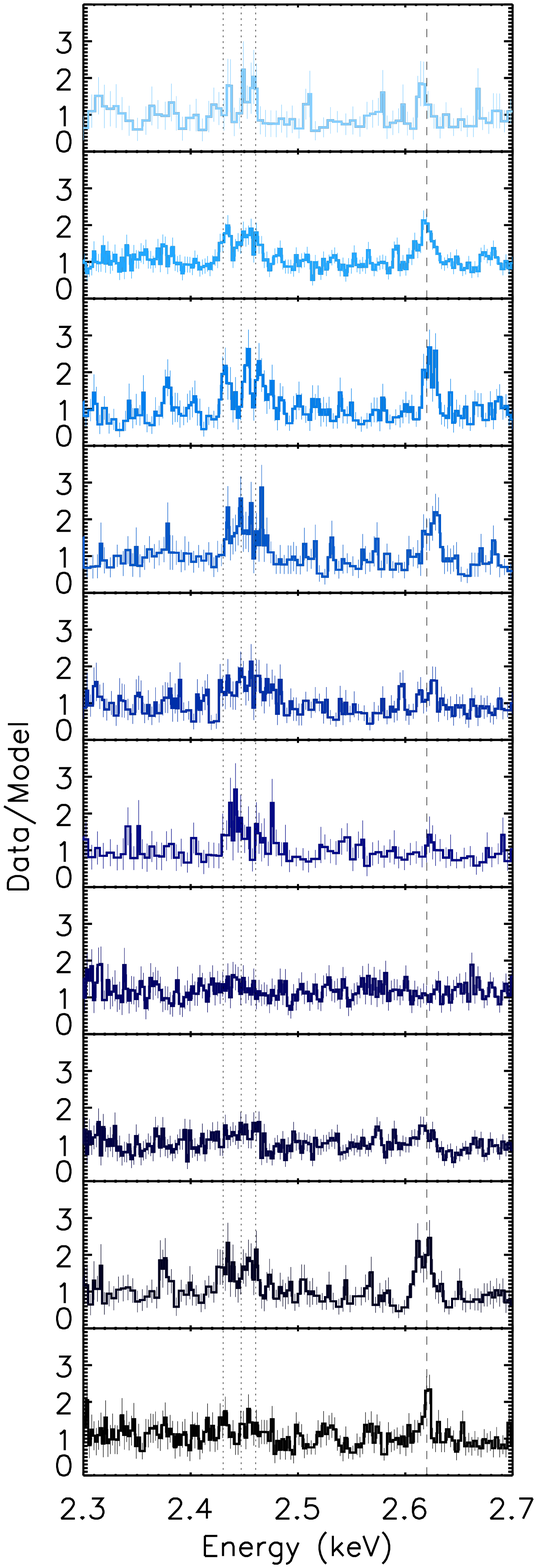}}
\subfigure[\label{fig:1}][]{\includegraphics[width=.23\linewidth,angle=0,clip=true,trim=0cm 2cm  0   6cm]{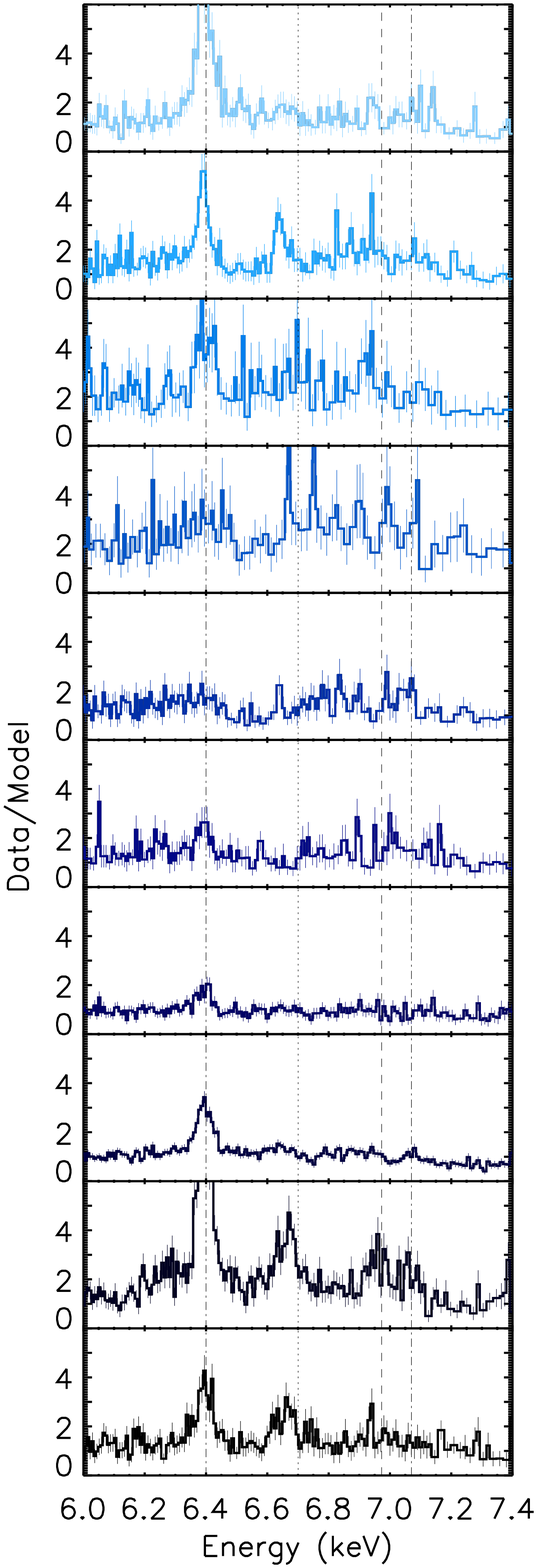}}

\subfigure[\label{fig:1}][]{\includegraphics[width=.23\linewidth,angle=0,clip=true,trim=1cm 2cm .5cm  10cm]{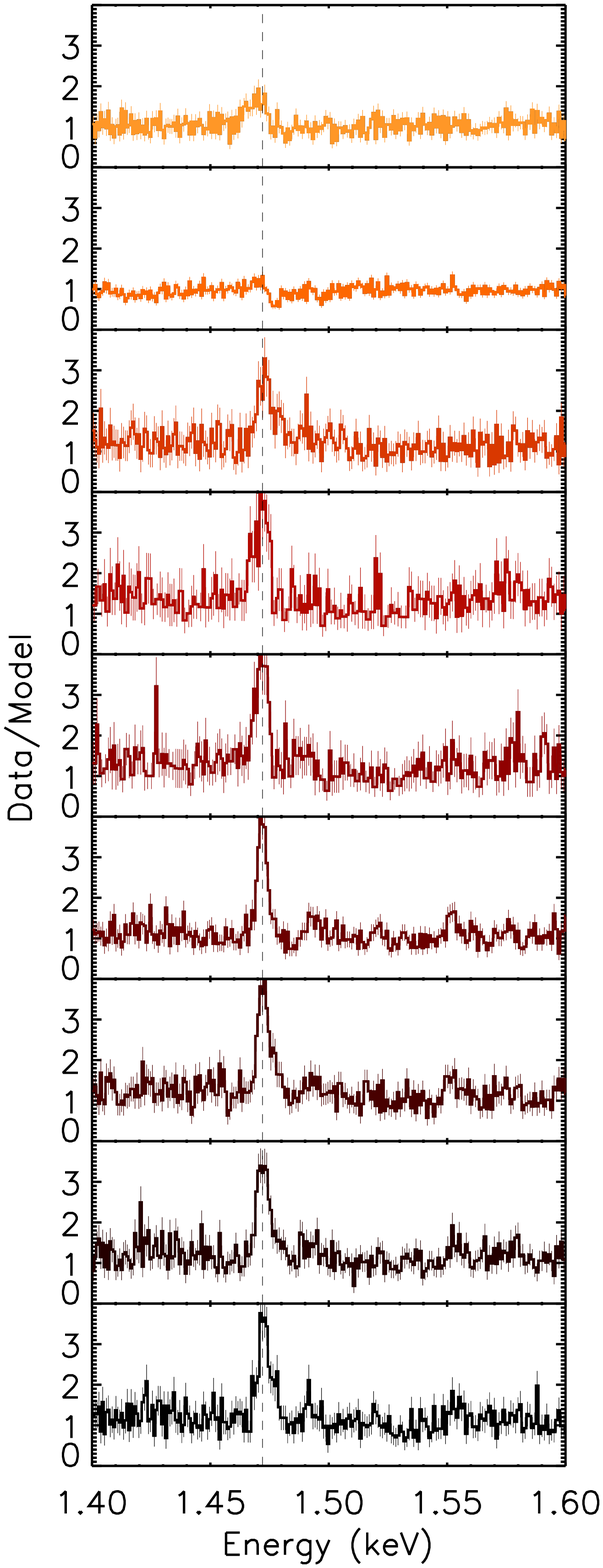}}
\subfigure[\label{fig:1}][]{\includegraphics[width=.23\linewidth,angle=0,clip=true,trim=1cm 2cm .5cm  10cm]{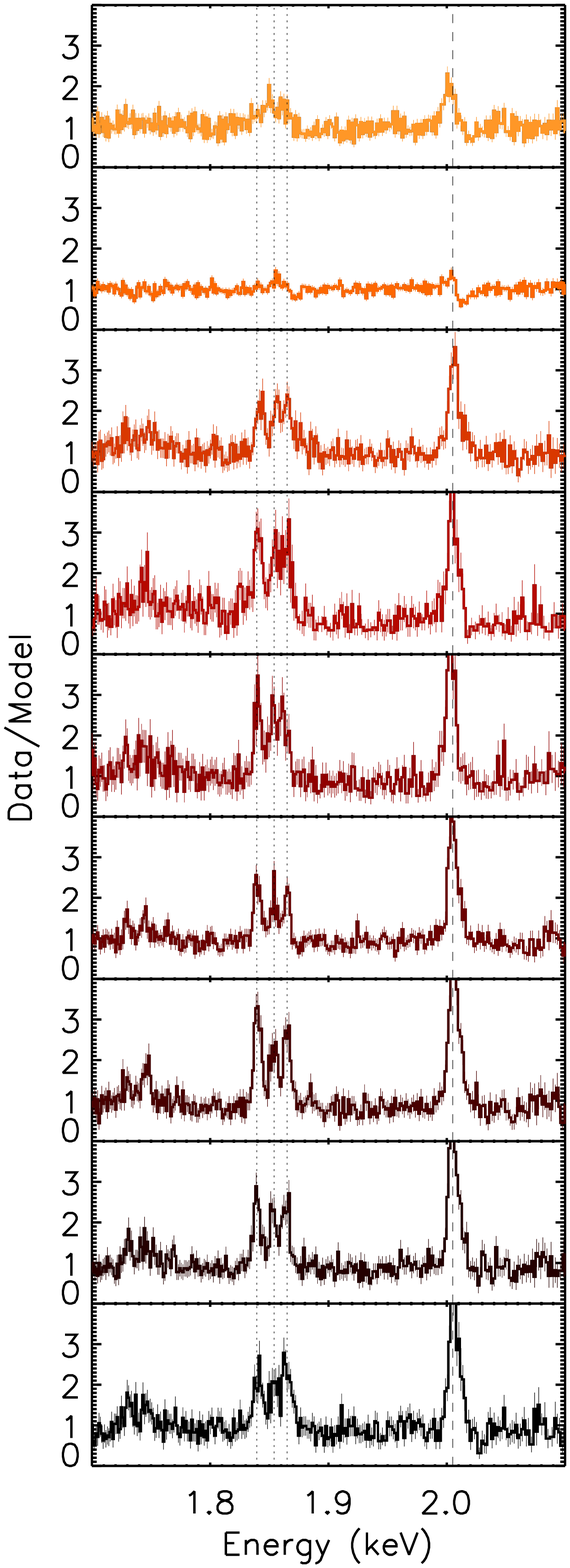}}
\subfigure[\label{fig:1}][]{\includegraphics[width=.23\linewidth,angle=0,clip=true,trim=1cm 2cm .5cm  10cm]{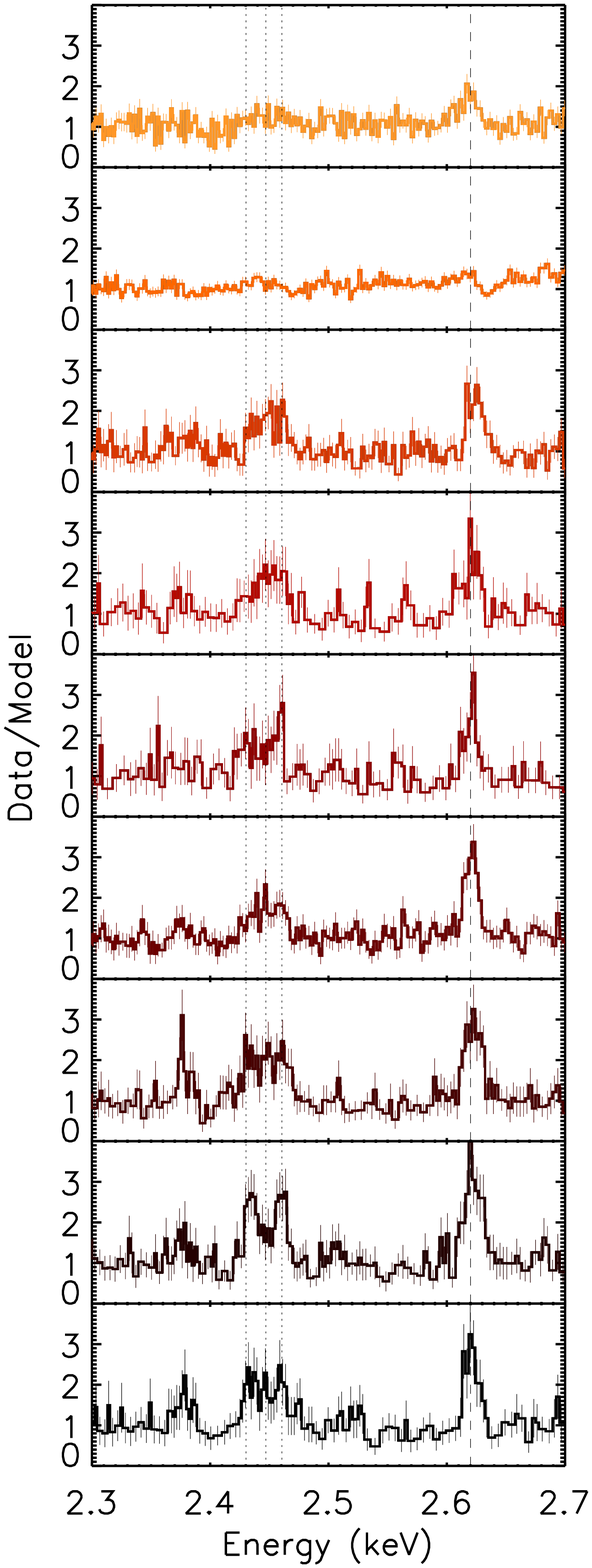}}
\subfigure[\label{fig:1}][]{\includegraphics[width=.23\linewidth,angle=0,clip=true,trim=1cm 2cm .5cm  10cm]{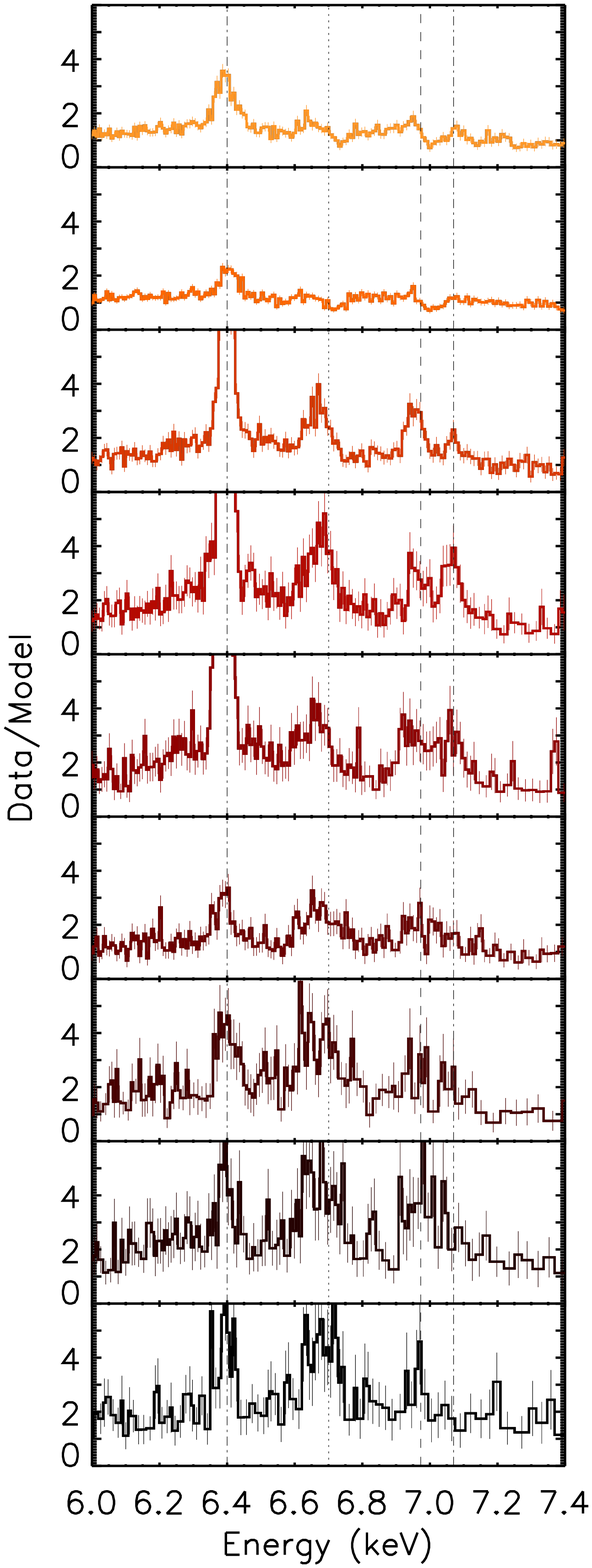}}
\caption{ These panels show the data to model ratio of each 3.2 ksec epoch in the first (a--d) and second (e--h) observations. Each segment is 3.2 kseconds, time proceeds from bottom to top. The dashed lines are Lyman-$\alpha$, dotted lines are He-like triplets, and the dot-dashed lines are the Fe K$\alpha$ and K$\beta$ lines. \label{fig:obs1} }
\end{figure*}

\begin{figure*}
\subfigure[\label{fig:1}][]{\includegraphics[width=.33\linewidth,angle=0]{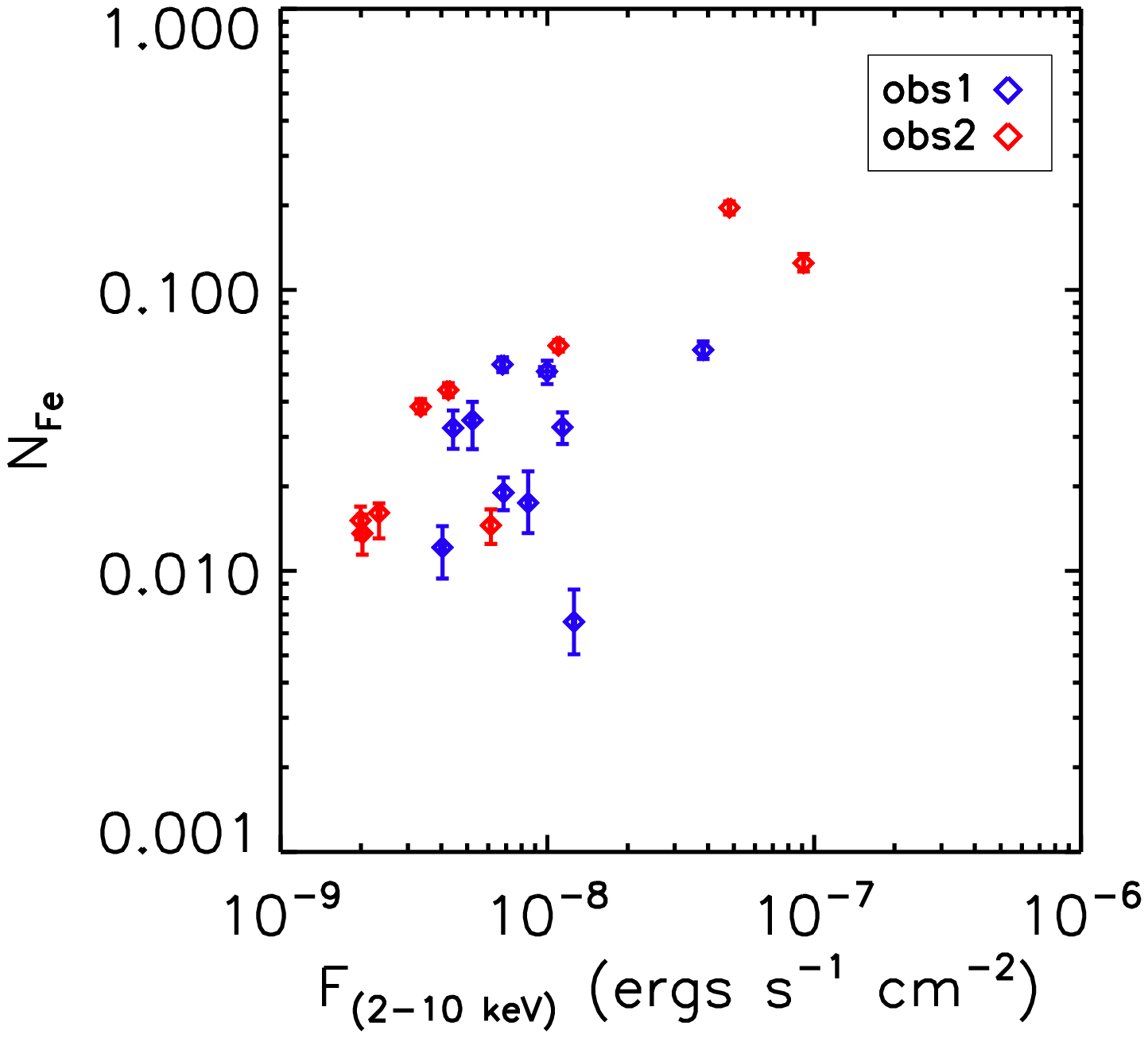}}
\subfigure[\label{fig:1}][]{\includegraphics[width=.33\linewidth,angle=0]{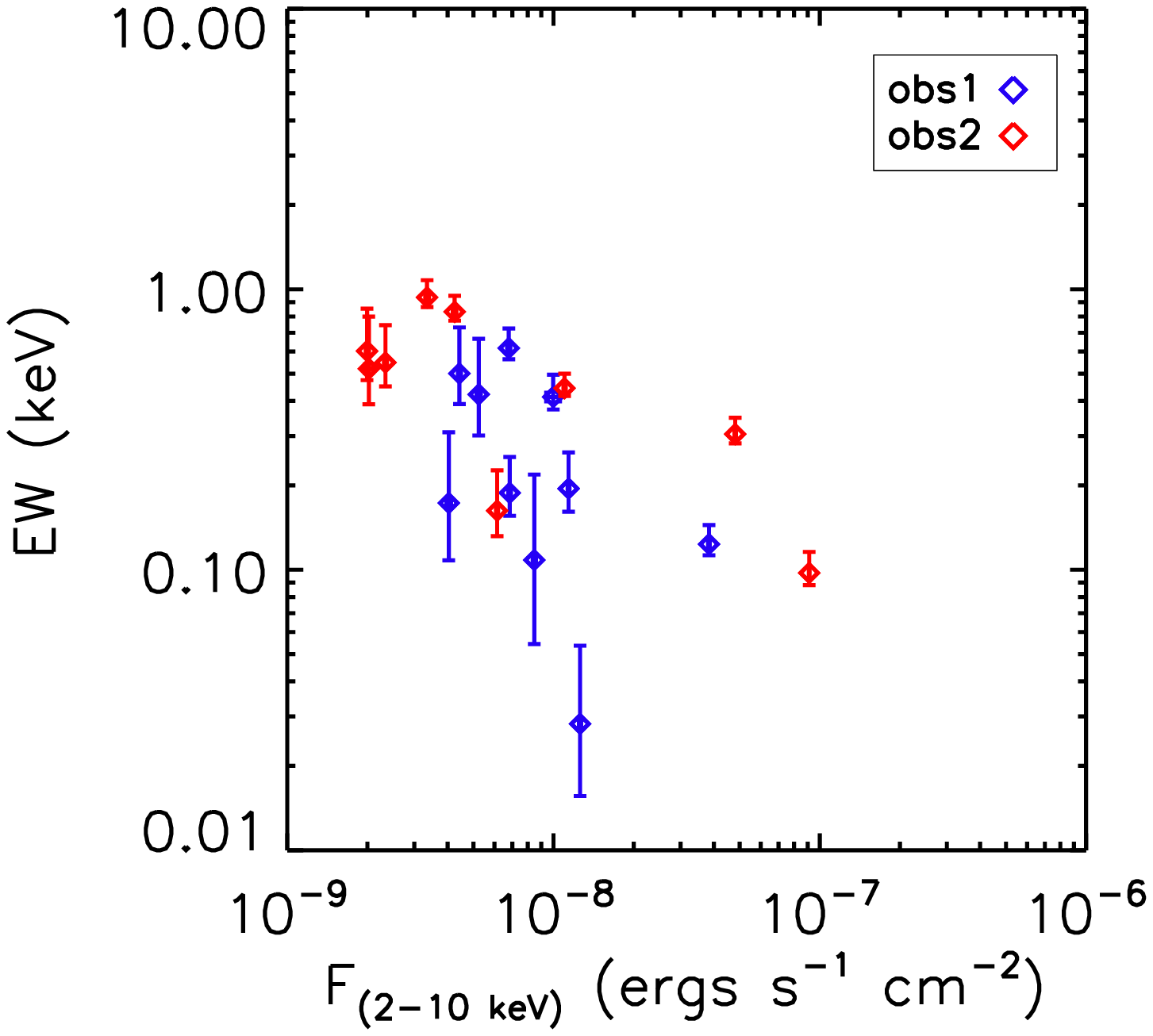}}
\subfigure[\label{fig:1}][]{\includegraphics[width=.33\linewidth,angle=0]{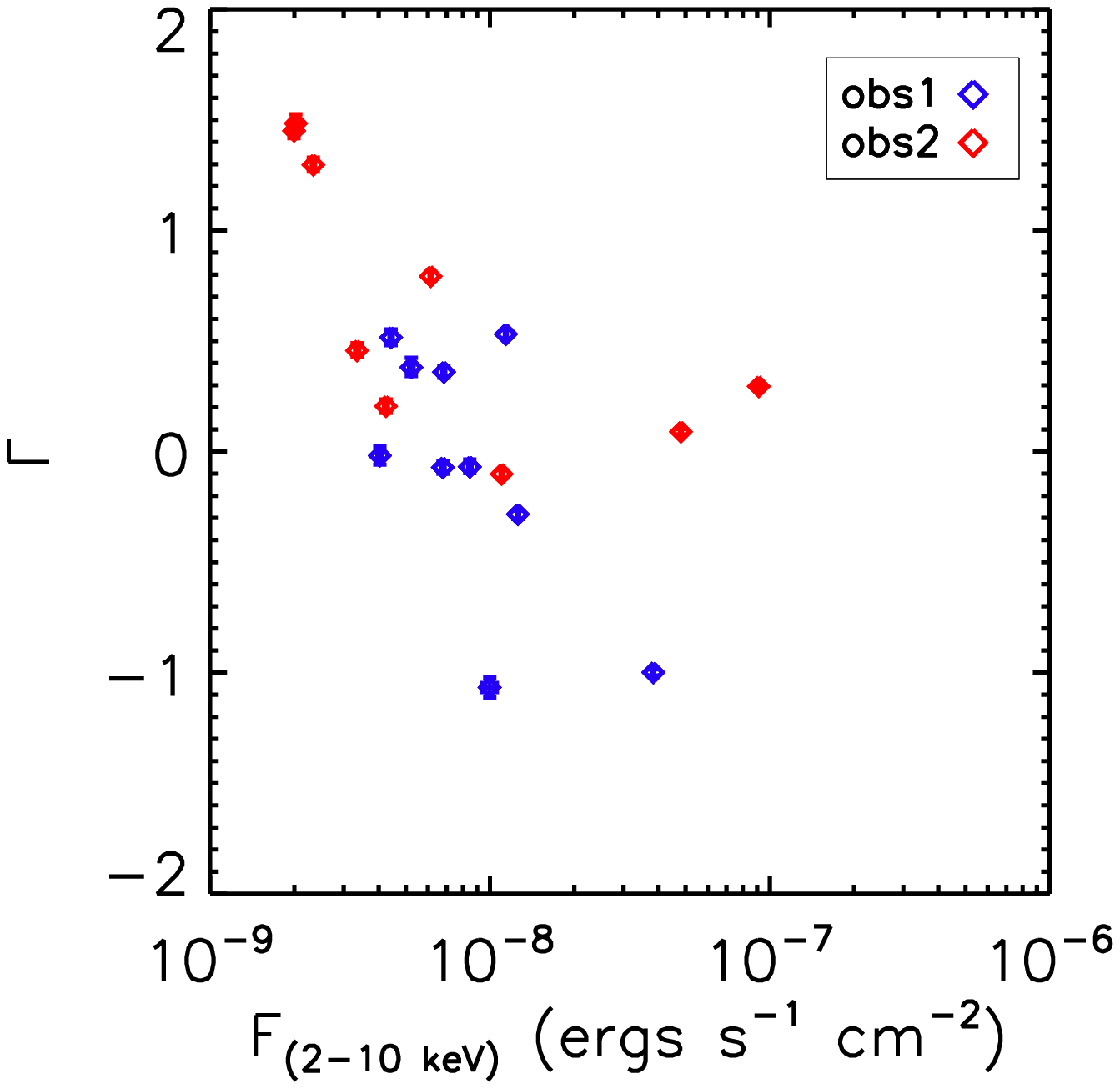}}

\caption{ Panel A shows the line intensity of the narrow Fe K$\alpha$ line to the total flux measured between 2--10 keV. A tentative positive trend is noted between these quantities. Panel B shows a negative correlation between the equivalent width of the narrow Fe K$\alpha$ and the total flux. The extremely high value of 1keV in the lower flux states suggest that we are not observing the direct continuum, but the outer disk may be obscuring our view. The resulting continuum is a combination of scattered and reflected light from the outer and inner disk, respectively. Panel C shows the phenomenological power-law spectral index versus the total flux. $\Gamma<1.4$ can not be produced by a simple Comptonization model, and therefore absorption, scattering and reflection from the disk are likely altering our view of the continuum.}
\end{figure*}

\begin{figure*}
\subfigure[\label{fig:1}][]{\includegraphics[width=.23\linewidth,angle=0,clip=true,trim=0cm 2cm .5cm  10cm]{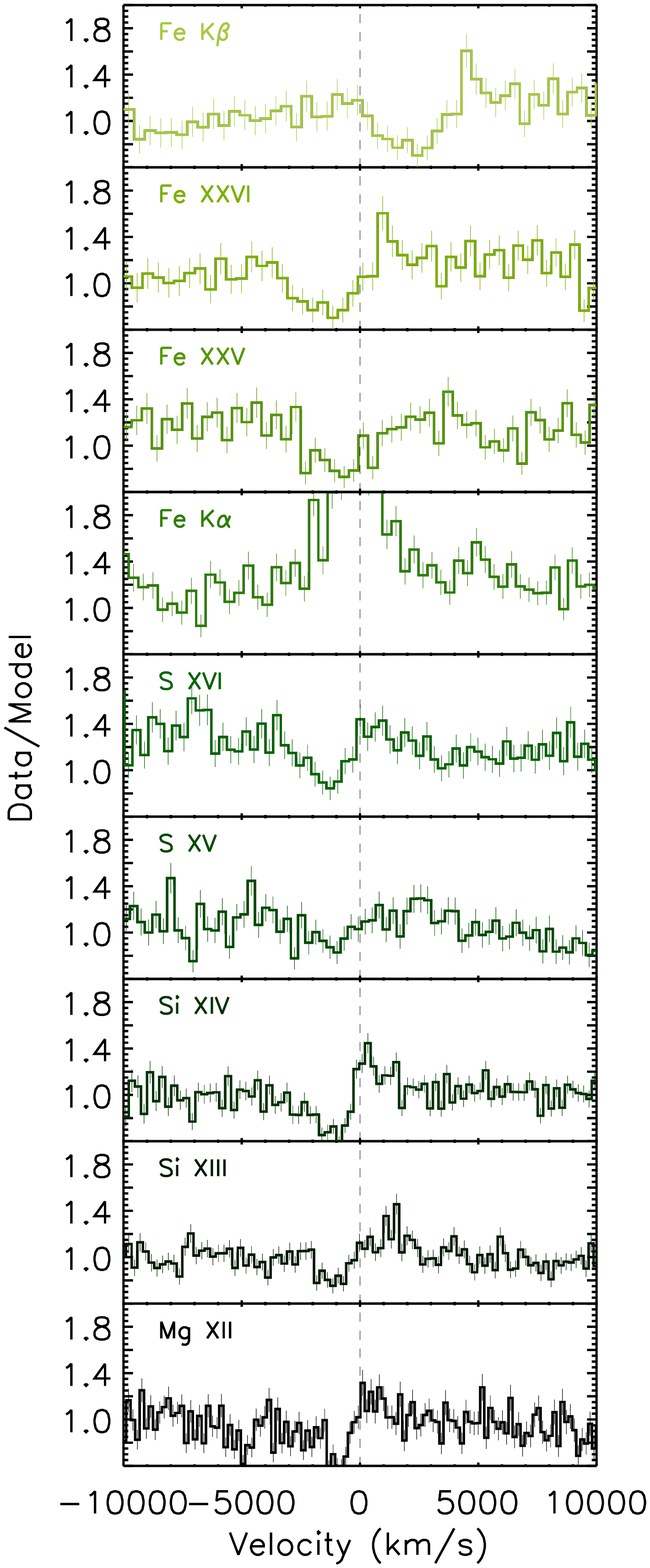}}
\subfigure[\label{fig:1}][]{\includegraphics[width=.23\linewidth,angle=0,clip=true,trim=0cm 2cm .5cm  10cm]{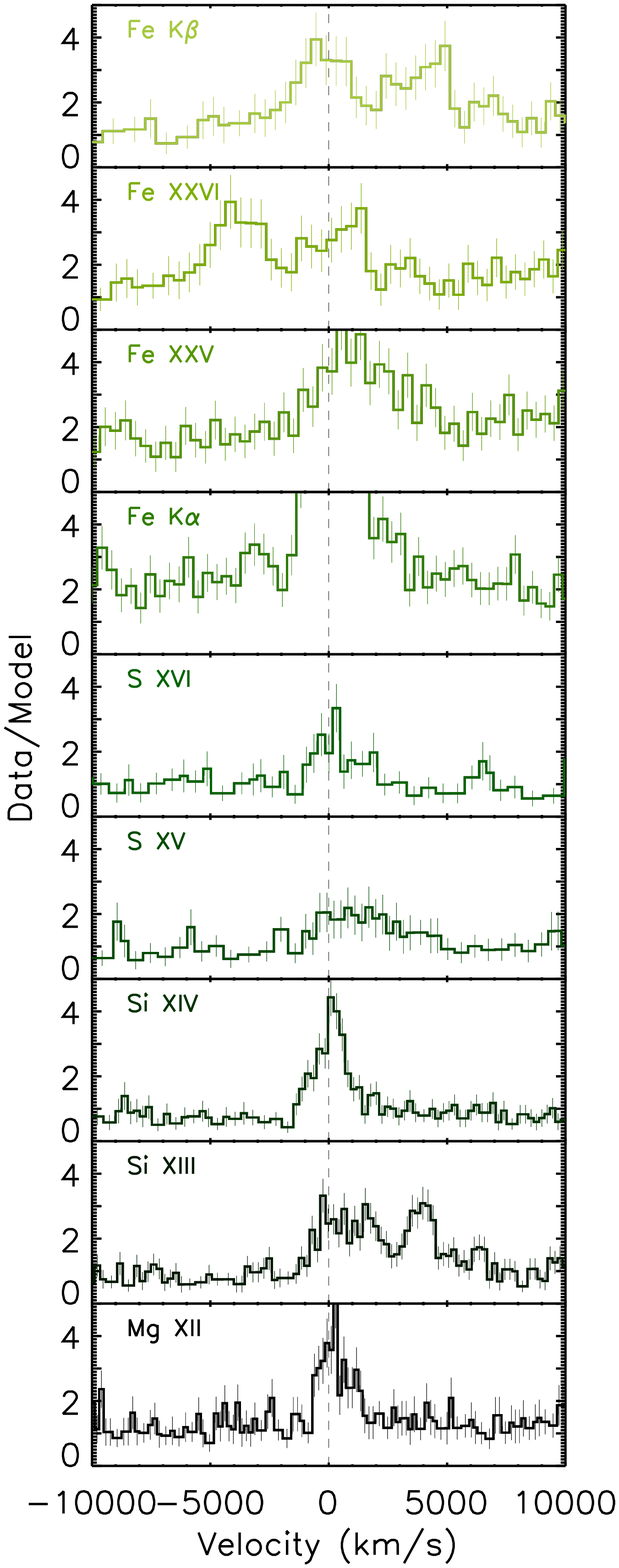}}

\caption{ Panel A corresponds to the 8th epoch in the second observation, and shows the ratio of the data to a phenomological power-law component. Clear absorption features are observed in all but the Fe K$\alpha$ and Fe K$\beta$ lines. The absorption line width increases from Mg XII to Fe XXVI.  For contrast, Panel B shows the same ions from the 6th epoch of the second observation. Only emission is detected in this epoch. \label{fig:obs1} }
\end{figure*}


\end{document}